\documentclass[reprint,amsmath,amssymb,aps,prapplied,superscriptaddress,bibnotes,floatfix]{revtex4-2}

\usepackage{graphicx}
\usepackage{dcolumn}
\usepackage{bm}
\usepackage{xcolor}
\usepackage{braket}

\usepackage[T1]{fontenc}
\usepackage{lmodern}
\usepackage{textcomp}
\usepackage{caption}
\usepackage{subcaption}

\begin{document}

\preprint{APS/123-QED}

\title{Temporally multiplexed ion-photon quantum interface via fast ion-chain transport
}

\author{Bingran You}\thanks{These authors contributed equally to this work.}
\affiliation{Department of Physics, University of California, Berkeley, CA 94720, USA}
\affiliation{Lawrence Berkeley National Laboratory, Berkeley, CA 94720, USA
}

\author{Qiming Wu}\thanks{These authors contributed equally to this work.}
\email{qiming.wu@berkeley.edu}

\affiliation{Department of Physics, University of California, Berkeley, CA 94720, USA}

\affiliation{Lawrence Berkeley National Laboratory, Berkeley, CA 94720, USA
}

\author{David Miron}
\affiliation{Department of Physics, University of California, Berkeley, CA 94720, USA}
\affiliation{Lawrence Berkeley National Laboratory, Berkeley, CA 94720, USA
}

\author{Wenjun Ke}
\affiliation{Department of Physics, University of California, Berkeley, CA 94720, USA}

\author{Inder Monga}
\affiliation{Lawrence Berkeley National Laboratory, Berkeley, CA 94720, USA
}

\author{Erhan Saglamyurek}
\affiliation{Department of Physics, University of California, Berkeley, CA 94720, USA}
\affiliation{Lawrence Berkeley National Laboratory, Berkeley, CA 94720, USA
}

\author{Hartmut Haeffner}
\email{hhaeffner@berkeley.edu}
\affiliation{Department of Physics, University of California, Berkeley, CA 94720, USA}
\affiliation{Lawrence Berkeley National Laboratory, Berkeley, CA 94720, USA
}

\def\thefootnote{*}\footnotetext{These authors contributed equally to this work}\def\thefootnote{\arabic{footnote}}

\date{\today}

\begin{abstract}
High-rate remote entanglement between photon and matter-based qubits is essential for distributed quantum information processing. A key technique to increase the modest entangling rates of existing long-distance quantum networking approaches is multiplexing. Here, we demonstrate a temporally multiplexed ion-photon interface via rapid transport of a chain of nine calcium ions across 74\,\textmu m within 86\,\textmu s. The non-classical nature of the multiplexed photons is verified by measuring the second-order correlation function with an average value of $g^{(2)}(0)$ = 0.060(13). This indicates low crosstalk of about 1\% between the multiplexed modes, which can be reduced further once coupling of the photons to a single-mode fiber is incorporated. In addition, we characterize the motional degree-of-freedom of the ion crystal after transport and find that it is coherently excited to $\bar{n}_\alpha\approx 110$ for the center-of-mass mode. Our proof-of-principle implementation paves the way for large-scale quantum networking with trapped ions, but highlights some challenges that must be overcome.

\end{abstract}

\maketitle

\section{Introduction}

Remote entanglement across distant quantum nodes~\cite{kimble2008quantum,duan2001long} may be used for long-distance quantum key distribution~\cite{van2022entangling,nadlinger2022experimental}, modular quantum computer architectures~\cite{monroe2014large,covey2023quantum}, as well as quantum enhanced metrology and sensing~\cite{komar2014quantum,nichol2022elementary,guo2020distributed}. Light-matter quantum interfaces are likely key building blocks for such applications as they allow for distributing entanglement between stationary matter qubits by using "flying" photons.

For practical purposes, these quantum interfaces need to be capable of establishing remote entanglement at high rates across a large-scale network of quantum nodes. However, in widely adopted probabilistic schemes based on heralded photon detection~\cite{moehring2007entanglement,hofmann2012heralded, bernien2013heralded,sangouard2009quantum}, the attempt rate is bounded by the round-trip travel-time of photons in a single mode, severely limiting the obtainable rates even at distances of a few kilometers. For example, for a communication distance of 100\,km, photon-travel time will be around 1\,ms, such that the attempt rate cannot be larger than $\mathrm{10^3}\,\text{s}^{-1}$. Even with the use of a state-of-the-art photon-matter interface that can yield close to unity photon extraction efficiency~\cite{schupp2021interface}, the long-distance matter-photon entanglement rate would be limited to sub 10\,$\text{s}^{-1}$ where we assumed a total loss of 20\,dB after considering the collection probability of generated photons, quantum frequency conversion efficiency, photon detector efficiency, and the optical fiber loss~\cite{krutyanskiy2023telecom,krutyanskiy2023multimode}.

A key solution to this challenge is multiplexing: to combine multiple signals into a single channel and therefore increase the attempt rate~\cite{pittman2002single,kaneda2015time}. Multiplexing has become a mature technology for ensemble-based quantum interfaces, both in atomic gases and rare-earth-ion-doped solid-state systems~\cite{pu2017experimental,sinclair2014spectral,saglamyurek2016multiplexed,zhang2024fast}. However, large-scale local quantum information processing is technically challenging in these platforms~\cite{bradley2019ten}. 

\begin{figure}[!t]
    \centering
    \includegraphics[width = 0.48\textwidth]{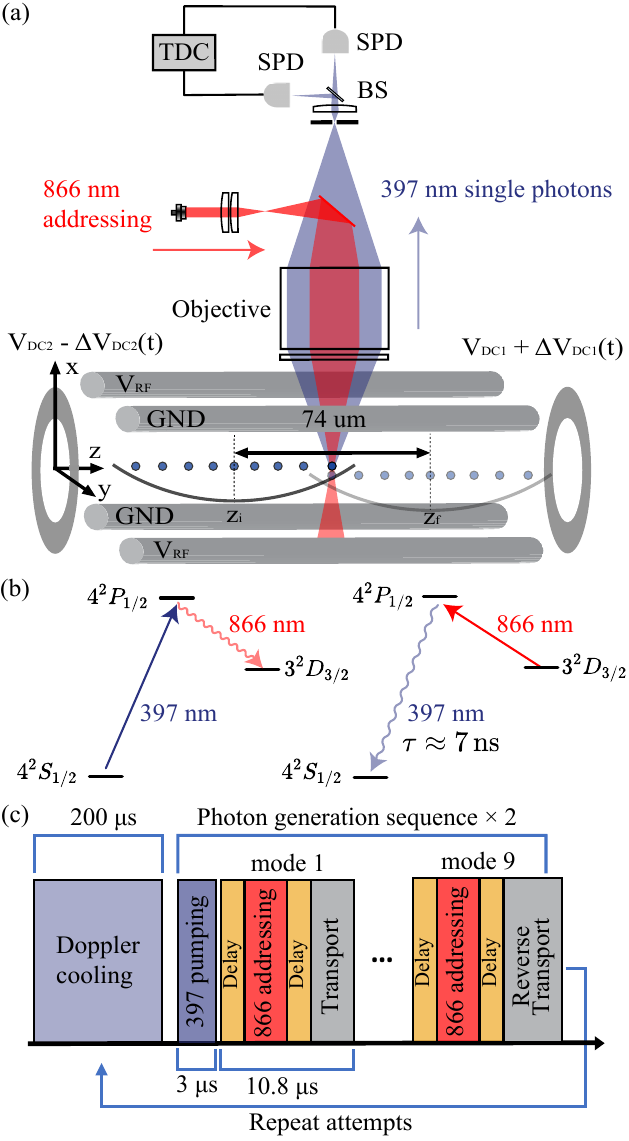}
    \caption{Schematics of multiplexed ion-photon interface. (a) A nine-ion chain is confined in an RF Paul trap. Controlling DC endcap voltages allows for ion transport. A beam of 397\,nm and 866\,nm light illuminating all ions is used for Doppler cooling. An objective collects the 397\,nm single photons and guides them to a 50/50 beamsplitter, followed by a photomultiplier tube on each exit port for photon detection. An 866\,nm beam counter-propagates with the single photons and addresses individual ions. (b), (c) Excitation scheme and pulse sequence for the 397\,nm single-photon generation. First, a global 397\,nm beam prepares the ions in the $3 ^2D_{3/2}$ state. Then, the 866\,nm addressing beam (resonant with $3 ^2D_{3/2}\leftrightarrow 4 ^2P_{1/2}$) is stroboscopically switched on when the target ion is in the interaction zone to extract photons from the target ion.}

    \label{fig:illustration}
\end{figure}

In contrast, single emitters, including trapped ions and neutral atoms, offer excellent local quantum information processing capability besides their natural interface with light at convenient wavelengths for quantum frequency conversion (QFC)~\cite{krutyanskiy2017polarisation,saha2023lownoise}, and the possibility of long-lived storage of entanglement~\cite{wang2021single,drmota2023robust}. On the other hand, implementing a multiplexed light-matter interface with these systems is technically challenging. Towards overcoming this problem, a few multiplexing schemes have already been proposed for ion- and atom-based quantum processors~\cite{huie2021multiplexed,li2024high,dhara2022multiplexed,ramette2022any}.
In view of the recent advances of the QCCD (quantum charge-coupled device) architecture~\cite{bowler2012coherent,pino2021demonstration,moses2023race}, a compatible approach to multiplexing is the process of ion-transport through a specific spatial location with maximized photon coupling efficiency.

In this work, we demonstrate such a temporal multiplexing scheme based on the transport of an ion-chain that can be used to improve the rate of ion-photon entanglement over long distances. In our experiments, we generate on-demand single photons by shuttling a nine-ion chain across the focus of a single-ion addressing beam. This scheme is expected to lead to a nearly nine-fold increase in attempt rate of the entanglement generation for quantum repeater nodes separated by >100\,km. We verify the single-photon nature of the photon trains by measuring a second-order time correlation of $g^{(2)}(0)$ = 0.060(13) without background subtraction. Furthermore, we investigate the problem of motional excitation during the transport, which is detrimental to local entangling operations~\cite{webb2018resilient} and, in the case of using a cavity for stimulating the photons, would lead to uncertainty in the coupling strength~\cite{takahashi2020strong}. Using a shuttling function designed to mitigate motional excitation, we find coherent excitation as high as $\bar{n}_\alpha\sim$~110 on the center-of-mass (COM) mode during one round of ion chain transport. These results show that the proposed multiplexing scheme can be scaled up to higher rates provided that more refined transport methods are applied.

\section{Multiplexed single-photon generation}

 The schematics of the experimental procedures are illustrated in Fig.\,\ref{fig:illustration}(a). Our experiment is conducted using an RF Paul trap similar to the one described in~\cite{broz2023test}. It is composed of four RF blades for generating the radial pseudopotential and two DC endcap electrodes for providing axial harmonic confinement. We typically trap a chain of nine $\mathrm{^{40}Ca^+}$ ions in a linear configuration with COM mode frequencies of $\omega_{x} = 2\pi\times 1.15$\,MHz in the radial and $\omega_{z} = 2\pi\times 0.179$\,MHz in the axial direction.
 Two global laser beams at 397\,nm and 866\,nm [not shown in Fig.\,\ref{fig:illustration}(a)] evenly illuminate the entire ion chain for Doppler cooling and optical pumping, and a tightly focused 866\,nm addressing beam with beam waist $\approx$ 6.0\,\textmu m allows for resonant excitation to extract the 397\,nm single photons from individual ions. The generated single photons are collected by an objective with a numerical aperture of NA\,=\,0.3 (collection efficiency $ P_{c}\approx2.5\%$) and directed to a 50/50 beam splitter (BS). At the exit port of the BS, photons are detected by two photomultiplier-tube (PMT)-based single-photon detectors (SPDs), and their arrival times are recorded with a time-to-digital converter (TDC) for subsequent analysis.

 We generate single photons on demand based on a background-free scheme, as illustrated in Fig.~\ref{fig:illustration}(b)~\cite{takahashi2013integrated}. In this process, we first perform Doppler cooling (detuning $\Delta = -\Gamma/2$) of the ion chain for 200\,\textmu s, with the first 100\,\textmu s assisted by another beam $-500$\,MHz detuned from the $4 ^2S_{1/2}\leftrightarrow 4 ^2P_{1/2}$ transition (not shown in Fig.\,\ref{fig:illustration}) to mitigate collision-induced ion chain melting in the Paul trap~\cite{van2022rf}. Then we begin the photon generation sequences with optical pumping to the $3^2D_{3/2}$ state (lifetime $\tau \approx 1.2$\,s) for 3\,\textmu s, followed by transport of the chain to position each ion in the focus of the 866\,nm addressing beam (NA = 0.1, waist 6\,\textmu m) resonant with the $3 ^2D_{3/2}\leftrightarrow 4^2P_{1/2}$ transition and to generate 397\,nm single photons [see Fig.~\ref{fig:illustration}(c)]. The ion transport and photon generation sequences are repeated twice between each Doppler cooling cycle.

\begin{figure}[!t]
    \centering
    \includegraphics[width = 0.48\textwidth]{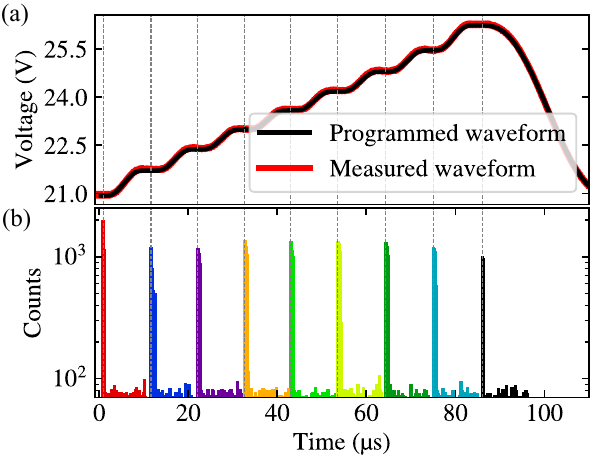}
    \caption{Temporal profile of single-photon generation. (a) The black (red) line is the programmed (measured) voltage ramp on endcap 1 during the forward transport, showing negligible latency effect. The voltage on endcap 2 is an inverse function of endcap 1 with an offset of +2.51\,V. (b) Detection time of the photons (bin width 16\,ns) color-coded with the associated ion (temporal mode).}

    \label{fig:photon_profile}
\end{figure}

\begin{figure}[!b]
    \centering
    \includegraphics[width = 0.5\textwidth]{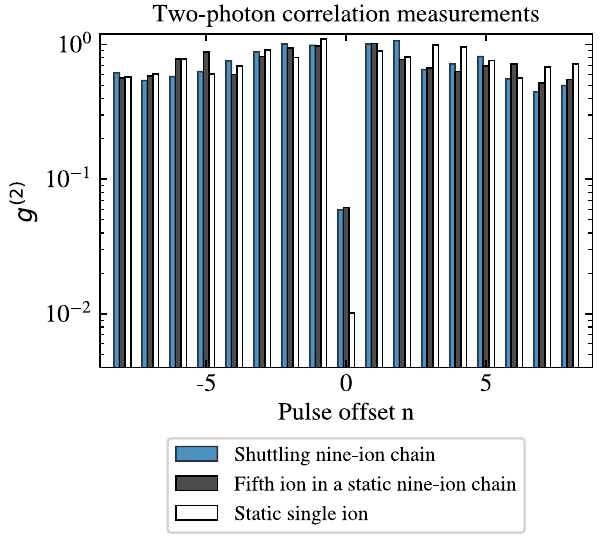}
    \caption{Two-photon correlation measurements. Blue bars represent $g^{(2)}$ measurements when shuttling a nine-ion chain; the average $g^{(2)}(0)$ in this case is found to be $0.060(13)$. The horizontal axis indicates delay time, quantified by the number of attempts in between. The shuttling function in Fig.~\ref{fig:photon_profile} is repeated twice before the next cooling cycle (see Fig.~\ref{fig:illustration}). Black bars represent $g^{(2)}$ measurements when addressing the fifth ion in a static nine-ion chain, $g^{(2)}(0)$ = 0.062(20). There are 25 attempts after each cooling cycle. White bars represent $g^{(2)}$ measurements of a single ion with the addressing beam using the same pulse sequence, $g^{(2)}(0)$ = 0.010(6).}
    \label{fig:g_2}
\end{figure}

During the ion-chain transport process, the endcap voltages are controlled by an arbitrary waveform generator (AWG) amplified by a custom-made, low-noise amplifier circuit with a gain of ten through low-pass filters and a vacuum feedthrough. The low-pass filters have cutoff frequencies of 1.9\,MHz to allow fast transport of the ion chain close to the speed of the COM mode frequency. The programmed and the measured waveforms show a negligible latency effect from the filters [Fig.~\ref{fig:photon_profile}(a)]. The forward shuttling function has eight steps, during each of which a different ion is placed in the focus of the addressing beam for 1.7\,\textmu s with the 866\,nm beam turned on simultaneously. After completing this sequence, we move the entire ion chain back to the original position in 35\,\textmu s using the same functional form as used above in one step. We use a sigmoid-like polynomial function such that the first- and second-order derivatives of the endcap voltages $V_{1,2}(t)$ vanish at the beginning and the end of the transport~\cite{tobalina2020invariant}

\begin{equation}
    \begin{array}{l} V_\mathrm{DC1}(t) = V_\mathrm{DC1}+\Delta V\left(10 \left(\frac{t}{T}\right)^3-15 \left(\frac{t}{T}\right)^4+6 \left(\frac{t}{T}\right)^5\right) \\ V_\mathrm{DC2}(t) = V_\mathrm{DC2}-\Delta V\left(10 \left(\frac{t}{T}\right)^3-15 \left(\frac{t}{T}\right)^4+6 \left(\frac{t}{T}\right)^5\right)\end{array},
    \label{Eqn:Shuttling_function}
\end{equation}
where $\Delta V$ is the voltage difference between the beginning and the end of a step, $t$ is the time after the end of the previous step, and $T = 9.1$\,\textmu s is the time of each transport step. The details of voltage optimization and numerical simulation of motional excitation can be found in Appendices~\ref{app:voltage} and~\ref{app:transport}. We reconstruct the temporal profile of 397\,nm photons during transport using the recorded arrival times of photons on the PMTs.

Fig~\ref{fig:photon_profile}(b) shows the emission from individual ions, or temporal modes. Data were accumulated for 40\,minutes, during which around $9.36 \times 10^7$ attempts were made on the whole chain, corresponding to an attempt rate of $39.0\times10^3\textrm{/s}$, single-photon count rate of  
187\,counts per second and an average photon extraction efficiency of 0.48\,\%. The total insertion loss to the detector accounts for 
the quantum efficiency of the detector $\sim 30\%$ (Hamamatsu H10682-210). Single-photon generation efficiency using the 866\,nm addressing beam is around $ 60\%$, limited by the weak excitation used to suppress neighboring-ion emission.

\section{Two-photon correlation}

Next, we perform a two-photon correlation experiment to test the non-classical characteristics of our multiplexed photon source~\cite{diedrich1987nonclassical}. The probability of two-photon correlation when detecting a correlation event on two detectors at different times is given by
\begin{equation}
    \rho_c(\tau) = \rho_1(\tau)  \rho_2(\tau+n\times\delta T),
\end{equation}
where $\rho_1(\tau)$ and $\rho_2(\tau+n\times\delta T)$ are the probabilities of detecting a photon at $t = \tau$ and $\tau+n\times\delta T$ on detectors 1 and 2. Here, $n\times\delta T$ represents the time delay between detection events separated by 
$n$ photon generation attempts (modes). We define the second-order correlation function $g^{(2)}(n\times\delta T)$ as the coincidence count rate between the two detectors with a delay of $n\times\delta T$ in the entire measurement time, normalized such that the maximum value among all $g^{(2)}(m\times\delta T)$, for $m\in\{n\}$, is equal to 1.
In Fig.~\ref{fig:g_2}, the blue bars show the normalized correlation counts as a function of the delay mode window. We choose a coincidence window of 300\,ns in each mode and measure 8 coincident counts at zero delay in 4.8 hours, corresponding to a $g^{(2)}(0)$ = 0.060(13) averaged over all ions.  The residual correlation can be explained by excitation of neighboring ions, i.e., crosstalk of the addressing beam that is separately characterized to be $0.99\,\%$ using fluorescence of the nine-ion chain on a camera, corresponding to expected average $g^{(2)}_{\text{exp}}(0) = 0.049(8)$ (see Appendix~\ref{app:crosstalk}). To verify this hypothesis, we compare these results with those of using only a single ion as well as to addressing only the fifth ion in a static nine-ion chain (see the white and black bars in Fig.\,\ref{fig:g_2}). Here in contrast to the multiplexing experiment, recooling is performed after every 25 single-photon generation attempts. The two experiments yield $g^{(2)}(0)$ = 0.010(6) and $g^{(2)}(0)$ = 0.062(20) with 6.0 and 4.8 hours of data accumulation, respectively. The raw data of the $g^{(2)}$ measurements can be found in Appendix~\ref{app:rawg2}.

With the 866\,nm addressing beam crosstalk characterization, we estimate the expected residual correlation counts at zero delay time, corresponding to $g^{(2)}(0)$ in the second-order correlation function. We assume that, for each trial with 866\,nm addressing light, the probability of getting one photon from the center ion is $\rho_{0}$, the probability of getting one photon from the two nearest neighbor ions is $\rho_{s1}$ and $\rho_{s2}$, and the combined contribution of limited collection efficiency and loss along the optical path is $\rho_{\text{loss}}$. For convenience, we use $\tilde{\rho} = \rho \cdot (1 - \rho_{\text{loss}})$. Then, for non-zero delay time, if all other noise sources are ignored, we have correlation probability
$$
\rho_c(n) = (\frac{\tilde{\rho}_{0} + \tilde{\rho}_{s1} + \tilde{\rho}_{s2}}{2})^2, \,\,\, n \neq 0
$$
in which there is a factor of 2 in the denominator coming from the BS. For zero delay time, we have
$$
\rho_c(0) = [2\, (\frac{\tilde{\rho}_{0}}{2}) \cdot (\frac{\tilde{\rho}_{s1} + \tilde{\rho}_{s2}}{2}) + 2\, (\frac{\tilde{\rho}_{s1}}{2}) \cdot (\frac{\tilde{\rho}_{s2}}{2})].
$$
Therefore in the limit of small crosstalk, we have
$$
g^{(2)}_{\text{exp}}(0) \approx 2\cdot \frac{\tilde{\rho}_{s1} + \tilde{\rho}_{s2}}{\tilde{\rho}_{0}} + g^{(2)}_{0}(0),
$$
in which $g^{(2)}_{0}(0) = 0.010(6)$ corresponds to the contribution of other noise sources other than crosstalk, e.g., detector dark counts and ambient light, that have been characterized by single ion correlation measurement. Therefore, we expect $g^{(2)}_{\text{exp}}(0) = 0.050(13)$, which is the average $g^{(2)}(0)$ value for nine ions as nine single photon sources. It agrees well with our measurement result of $g^{(2)}(0) = 0.060(13)$ in the multiplexed single-photon generation.

While the single ion $g^{(2)}$ is limited by the detector dark counts and ambient light, the measured $g^{(2)}(0)$ of the static nine-ion chain appears to be limited by the crosstalk of the addressing beam. The results indicate the major source of residual correlation is addressing crosstalk (Appendix~\ref{app:crosstalk}). Because the ion-ion spacing is densest in the center ion, we would expect the average $g^{(2)}(0)$ of the moving chain to be lower than that of the static case where the fifth ion was addressed. We conjecture that the additional error is caused by the addressing beam becoming misaligned with the target ion due to lag in endcap voltage ramps or miscalibration. Finally, we note that similar to what is reported in~\cite{herold2016universal}, this crosstalk can be suppressed to <$10^{-5}$ by coupling the single photons into a single-mode fiber~\cite{o2024fast} or improving the optical quality of the addressing beam.

\begin{figure*}[t]
    \centering
    \includegraphics[width = \textwidth]{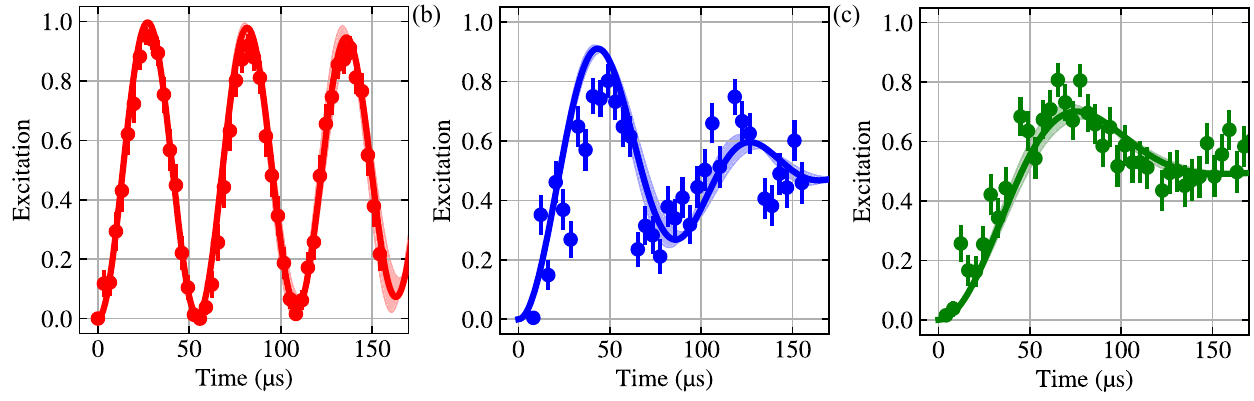}
    \caption{$\ket{\downarrow}\leftrightarrow\ket{\uparrow}$ carrier excitation of nine-ion chain before and after shuttling. The horizontal axis is the probe time of the global 729\,nm beam, and the vertical axis is the average ion excitation on the $\ket{\uparrow}$ state. Error bars denote one standard deviation of the quantum projection noise. (a) Rabi oscillations of the sideband-cooled ions (red dots).  The red line is a numerical simulation of thermal distribution with $\bar{n}_{th} = 4\pm 3$. (b) Rabi oscillation after the transport at half speed of the transport function in Fig.~\ref{fig:photon_profile}(a). The blue curve is a numerical simulation with $\bar{n}_{th}$ = 4 and a coherent excitation characterized by $\bar{n}_{\alpha}$ = $50\pm 5$. (c) Rabi oscillation after the transport at full speed. The green curve is a numerical simulation with $\bar{n}_{th}$ = 4, $\bar{n}_{\alpha}$ = $110\pm 5$.}

    \label{fig:heating_measurement}
\end{figure*}

\section{Motional excitation during transport}

Although the single-photon results demonstrate that the shuttling scheme implemented in this work is effective for stable single-photon generation, it is essential to investigate the motional excitation induced by ion transport as subsequent quantum operations on the atomic states require low motional excitation.
To characterize the motional state, we drive motion-sensitive Rabi oscillations on the $\ket{\downarrow}=\ket{4^2 S_{1/2},m_J=-1/2}\leftrightarrow\ket{\uparrow}=\ket{3^2 D_{5/2},m_J = -1/2}$ qubit transition using a narrow-linewidth 729\,nm beam on the ion chain, and use a PMT to record the average spin excitation. The differential force on the ions from the voltages transporting the chain is small and hence we expect that the transport excites mostly the COM mode to a coherent state.
Then, the average excitation can be written as
\begin{equation}
    P_e(t)=\frac{1}{2N}\left[1-\sum_{n=0}^{\infty}\sum_{i=0}^{N} P_n \cos \left(\Omega_{n}^{(i)}t\right)\right],
\end{equation}
where $P_n$ is the occupation on the $n$th number state and encodes a convolution between a thermal and a coherent phonon distribution~\cite{walther2012controlling}. $\Omega_{n}^{(i)}$ is the Rabi frequency of the $i$th ion on the $n$th number state, considering the Gaussian beam so that the intensity on each ion is slightly different.
Fig.~\ref{fig:heating_measurement}(a) shows the sideband-cooled carrier Rabi flopping before the ion transport with a fitted thermal distribution $\bar{n}_{th} = 4.0\,\pm\,3.0$. (b) shows the Rabi flopping after ion transport twice as slow as in Fig.~\ref{fig:photon_profile}(c). The blue curve corresponds to a numerical simulation using a convolution of the initial thermal population (fixed at $\bar{n}_{th}=4$) and a coherent excitation of the COM mode with $\bar{n}_\alpha = |\alpha|^2 \approx 50$. Similarly, for the full-speed transport, the Rabi flopping data are fit by a convolution of $\bar{n}_{th}=4$ and a coherent COM excitation of $\bar{n}_\alpha\approx 110$ (Fig.~\ref{fig:heating_measurement}(c)).

\begin{figure}[!hb]
    \centering
    \includegraphics[width = 0.48\textwidth]{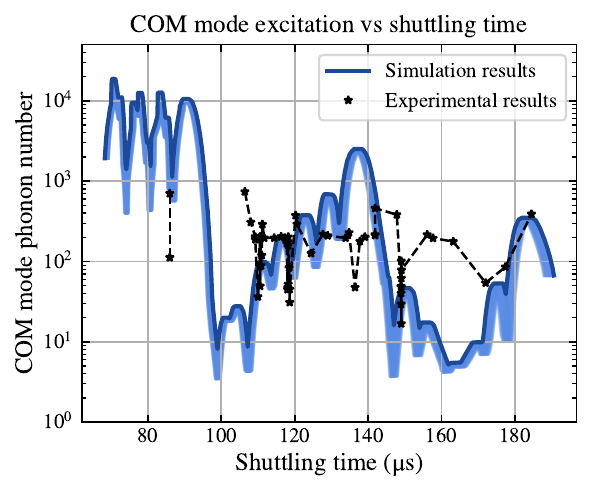}
    \captionsetup{justification=raggedright,singlelinecheck=false}
    \caption{Experimental measurements and numerical simulation of the COM-mode coherent excitation at the end of the shuttling process. Star-shaped markers represent the experimental data obtained at various shuttling durations. The blue curve indicates the calculated coherent excitation as a function of shuttling time under the trap frequency of $\omega_\mathrm{COM}/(2\pi) = 179(1)$\,kHz. The shaded region denotes the $\pm\sigma$ deviation corresponding to the measured trap frequency standard deviation of 1.01\,kHz throughout the shuttling region.}
    \label{fig:Simulation of COM}
\end{figure}

To better understand the coherent excitation during the ion transport, we experimentally measure and numerically simulate the phonon number excitation on the COM mode as a function of the total transport time (Fig.~\ref{fig:Simulation of COM}). The star-shaped markers depict the measured average phonon number on this mode after transport at various total shuttling times. The blue shaded region shows the numerical simulation using Eq.~\ref{eq:ODE}. The simulation results take into account the variation of $\pm\sigma$ axial trap frequencies due to spatial inhomogeneity as well as the measured voltage noise introduced by the arbitrary waveform generator. The solid line represents the simulation curve that considers the average of the above experimental imperfections, indicating that the phonon number is very sensitive to variation of the trap frequency and the total transport time.
However, the simulation does not fully match the experimental data. In particular, in the shuttling-time range of 160–170 $\mathrm{\mu s}$ the model predicts motional excitation below ten quanta that is significantly lower than the measured excitation values.
As our model does not involve the full treatment of diabatic long-chain transport, a comprehensive investigation is necessary to reproduce the observed behaviors in simulations. Future studies should include a more detailed model that accounts for axial–radial mode mixing during transport (leading to strong excitation of higher-order modes), the finite motional coherence time (preventing complete de-excitation of the motional energy), and an experimental characterization of the excitation of individual modes. Our detailed simulation and modeling are shown in Appendix~\ref{app:transport}. 

Optimal fast transport of a long ion chain is a complex problem, which we leave for future investigation. For example, further optimization can be explored by energy self-neutral shuttling~\cite{bowler2012coherent,walther2012controlling}, implementing closed-loop optimization of the shuttling function~\cite{sterk2022closed}, etc.

\section{Discussion and outlook}

To summarize, we have presented a multiplexed ion-photon interface by transporting a nine-ion chain with synchronized photon generation in sub-100\,\textmu s. We expect that the transport speed is limited by the axial trap frequency. Higher radial confinement allows for increased axial frequencies while preserving a linear ion chain of the same size, thereby enabling faster ion transport along the axial direction. 3D-printed micro traps have demonstrated an order of magnitude higher radial trap frequency\,\cite{Xu-2025-3D-traps}, thereby opening the path towards speeding up the interface by an order of magnitude. The 397\,nm photon can be converted to the telecommunication band via QFC~\cite{liu2026long,kasture2016frequency,saha2023lownoise}. Once integrated with state preparation on a $3 ^2D_{3/2}$ Zeeman sublevel and photon collection with a single-mode fiber, we expect a faster photon extraction rate~\cite{crocker2019high} and negligible ion crosstalk while achieving high-fidelity ion-photon entanglement~\cite{stephenson2020high,o2024fast}. Our approach can also be combined with a miniature cavity~\cite{takahashi2020strong} for much higher photon extraction efficiency without sacrificing the photon generation rate, while the 
positional spread of the ions caused by coherent excitation can be mitigated by aligning the cavity along the radial direction or further optimization of the shuttling function. These results stimulate research on fast shuttling of a chain of tens of ions as a unit cell of a logical qubit with heralded entanglement~\cite{li2024high} and high-rate entanglement of quantum processors across large distances.

During the review process of the manuscript, we became aware of related work on atom-transport-based multiplexing schemes for scalable quantum networks~\cite{canteri2024photon,cui2025metropolitan}.

\begin{acknowledgments}
We thank You-Wei Cheah, Ben Lanyon, Jiarui Liu, Tracy Northup, Alp Sipahigil, and Wenji Wu for their helpful discussions. Additionally, we are grateful to Dietrich Leibfried for valuable comments on diabatic ion chain transport. Q.W. and H.H. acknowledge funding by the U.S. Department of Energy, Office of Science, Office of Basic Energy Sciences under Award No. DE-SC0023277. This work is supported by the Office of Science (SC) in Advanced Scientific Computing Research (ASCR) through FOA—Quantum Internet to Accelerate Scientific Discovery (LAB 21-2495) and by NSF Grant No. 2016245.
\end{acknowledgments}

\appendix
\renewcommand{\thefigure}{A\arabic{figure}}
\setcounter{figure}{0}
\renewcommand{\theequation}{A\arabic{equation}}
\setcounter{equation}{0}

\section{Crosstalk of addressing laser}
\label{app:crosstalk}

In this section, we discuss the characterization of the intensity crosstalk of the 866\,nm addressing beam and corresponding correlation counts at zero delay time of our multiplexed photon source. 

Single-ion addressing technique with 729\,nm light has been demonstrated for quantum information processing~\cite{PhysRevA.60.145}. In this experiment, we use the objective that is optimized for 729\,nm (S6ASS2241, Sill Optics) to focus the 866\,nm light to generate single photons from individual ions. To quantify the intensity crosstalk of the nine-ion chain, we position each ion at the center of the addressing beam, optimize the power of the beam, and record the fluorescence of individual ions using an Electron Multiplying CCD (EMCCD) camera. The correlation between photon scattering rate $R_s$ and laser power $P$ for a two-level system is

\begin{equation}
        R_{s} = \gamma \cdot \frac{s}{2(1+s)},
\label{Eq:saturation}
\end{equation}

where $s = k \cdot P = \frac{P}{P_{s} \cdot (1+(2 \delta / \gamma)^2)}$ is the saturation parameter, $P_{s}$ is the saturation power, $\delta$ is the laser detuning from atomic resonance, and $\gamma$ is the decay rate of the excited state~\cite{van2016atoms}. With 397\,nm laser power and detuning fixed, we use this function to co-fit the fluorescence data of center ions and neighbor ions as a function of 866\,nm addressing laser power. Therefore, the intensity crosstalk equals $k_n / k_1$, with $k_1$ being the fitted parameter for the center ions and $k_n$ being the fitted parameter for an $n$th nearest-neighbor ion.

\begin{figure}[!hb]
    \centering
    \includegraphics[width = 0.48\textwidth]{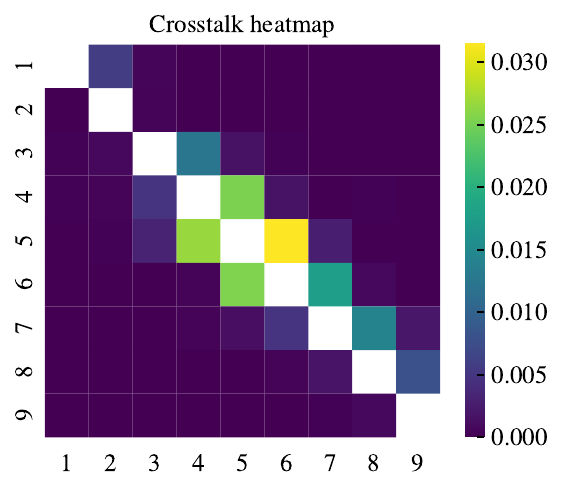}
    \captionsetup{justification=raggedright,singlelinecheck=false}
    \caption{Heatmap of crosstalk measurement of a nine-ion chain. The average nearest-neighbor intensity crosstalk is around 0.99\,\%.}
    \label{fig:heatmap_of_crosstalk}
\end{figure}

We characterize the crosstalk when the beam is centered on each ion. Fig.~\ref{fig:heatmap_of_crosstalk} illustrates the ratio of laser intensity on the non-addressed (neighbor) ions to the addressed (center) ion of the nine-ion chain. The average intensity crosstalk between the center and the nearest-neighbor ions is calculated to be 0.99\,\%. We note that the crosstalk is higher on one side of the addressing beam, which could be due to the positional mismatch of the ion chain relative to our addressing beam, or residual aberrations within the addressing beam itself.

\begin{figure}[!ht]
    \centering
    \includegraphics[width = 0.48\textwidth]{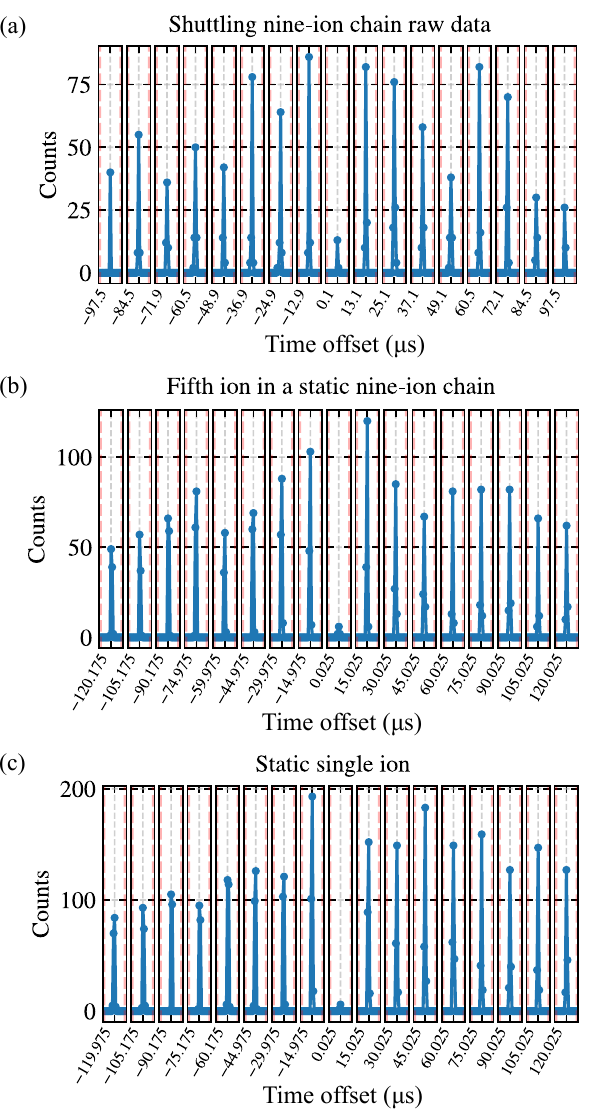}
    \caption{Raw experimental data for $g^{(2)}$ measurement with 200\,ns time bin for (a) shuttling a nine-ion chain (b) fifth ion in a static nine-ion chain and (c) a static single ion.}
    \label{fig:g_2_raw_data}
\end{figure}

\section{Raw experimental data for the two-photon correlation measurements}
\label{app:rawg2}

Figure~3 in the main text presents a bar-like plot comparing the $g^{(2)}$ measurements across three experimental scenarios: shuttling a nine-ion chain, addressing the fifth ion in a stationary nine-ion chain, and a single static ion. In that figure, the coincidence data are displayed as a function of relative pulse offsets (modes) rather than absolute detection times. That representation accounts for the different timescales of photon generation sequences among the experiments. For example, in the static ion case, the repetition rate is higher, with shorter time intervals between modes. However, this approach omits certain temporal details of coincidence measurement, such as the linewidth of the coincidence peak, which are informative about the photon wave packet. To provide this additional context, we present the raw experimental data from Fig.~3 in Fig.~\ref{fig:g_2_raw_data} using a time bin of 200\,ns.

\section{Voltage control of ion positions}
\label{app:voltage}

In this section, we illustrate the voltage levels for ion position control and numerical simulation of the shuttling process. Based on the simulation, the center of mass mode has the dominant motional excitation during the ion transport. We then show the experimental characterization of the transport-induced motional excitation as a function of transport time and compare it with numerical simulation.

\begin{figure}[!hb]
    \centering
    \includegraphics[width = 0.48\textwidth]{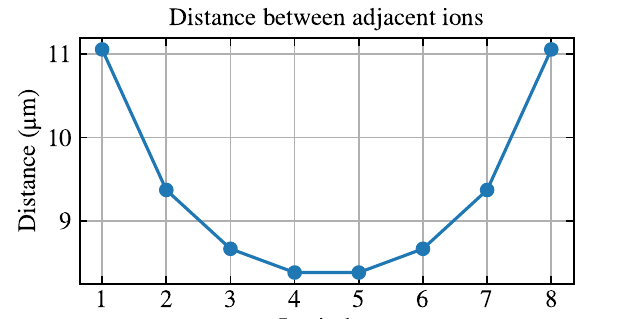}
    \captionsetup{justification=raggedright,singlelinecheck=false}
    \caption{Nearest-neighbor ion spacing of a nine-ion chain at an axial trap frequency of 179\,kHz, with average endcap voltage 24.24\,V.}
    \label{fig:equilibrium_positions}
\end{figure}

Control of ion positions during fast shuttling is crucial since misaligned ion positions may lead to insufficient photon extraction and increased crosstalk. To precisely align each ion directly at the center of the addressing beam for every voltage step, it is important to understand the ion-ion spacings in the chain. Then, we need to make sure the change in voltage levels matches the corresponding spacing between the ions. In this experiment the two DC endcaps provide the axial confinement. With average voltages of around 24.24\,V on the endcaps, the axial trap frequency is around 179\,kHz. Combining the Coulomb force and the harmonic potential, the relative equilibrium positions of ions are shown in Fig.~\ref{fig:equilibrium_positions}.

\begin{figure}[!ht]
    \centering
    \includegraphics[width = 0.48\textwidth]{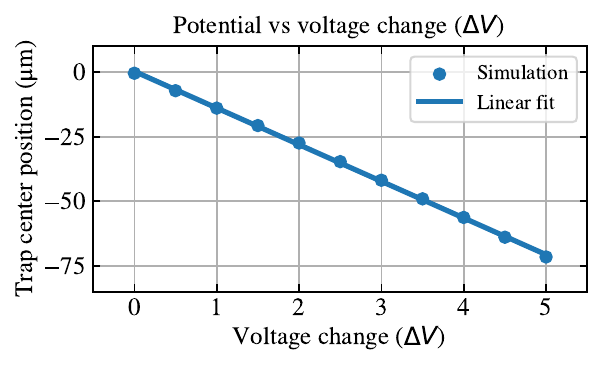}
    \captionsetup{justification=raggedright,singlelinecheck=false}
    \caption{Relation between axial trap center and endcap voltage change $\Delta {V}({t})$ with average endcap voltage 24.24\,V.}
    \label{fig:trap_frequency_and_average_endcap_voltage}
\end{figure}

To facilitate the transport of ion chains from one end to the other, we modulate the voltages by increasing voltage on one endcap and decreasing voltage on another while the average voltage remains unchanged. Fig.~\ref{fig:trap_frequency_and_average_endcap_voltage} shows the axial displacement of the trap potential's center as a function of voltage change on both endcaps, indicating a near linear relationship between the trap center position and $\Delta {V}({t})$. Therefore, we program the voltage level changes for each step referring to the ion-ion spacing, as shown in Fig.~\ref{fig:equilibrium_positions}. Each voltage ramp follows the function form with zero first- and second-order derivatives at the beginning and end (Eq.~1 in the main text~\cite{tobalina2020invariant}).

\section{Numerical simulation of shuttling process}
\label{app:transport}

\begin{figure}[!ht]
    \centering
    \includegraphics[width = 0.48\textwidth]{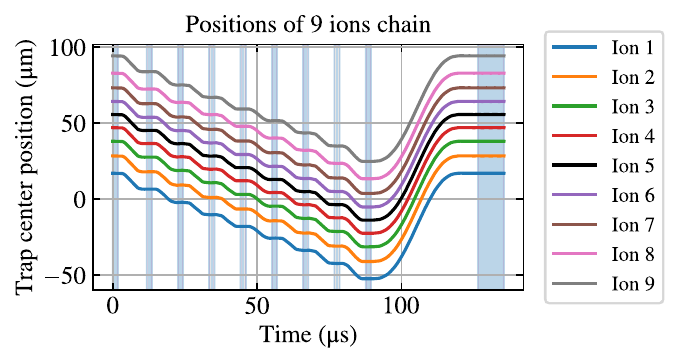}
    \captionsetup{justification=raggedright,singlelinecheck=false}
    \caption{The positions of the nine-ion chain as functions of time when performing half-speed shuttling. The first nine narrow blue areas are the time windows where voltage levels stay static. The last blue window is used to measure the motional excitation of all nine modes at the end of shuttling.}
    \label{fig:Shuttling function and ions positions}
\end{figure}

\begin{figure}[!b]
    \centering
    \includegraphics[width = 0.48\textwidth]{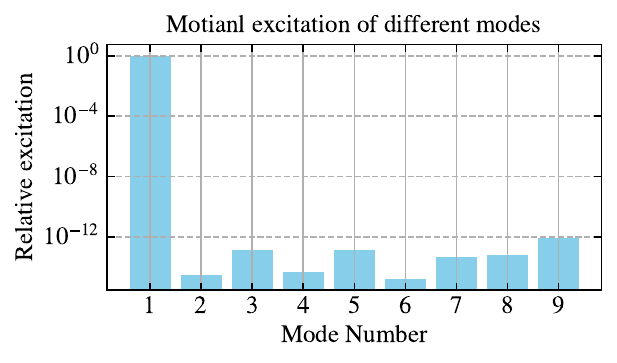}
    \captionsetup{justification=raggedright,singlelinecheck=false}
    \caption{Calculated excitation of nine motional modes along the axial direction at the end of the half-speed shuttling process. The COM mode has the dominant motional excitation. The mode excitations are normalized so that the COM mode excitation equals one.}
    \label{fig:Relative amplitudes of 9 motional modes along axial direction}
\end{figure}

The nine-ion chain transport can be described by ordinary differential equations (ODEs) as
\begin{equation}
\frac{d^2 z_i}{d t^2}=-\omega_0^2 z_i+\frac{k q^2}{m} \sum_{\substack{j=1 \\ j \neq i}}^N \frac{1}{\left(z_i-z_j\right)^2} \cdot \operatorname{sgn}\left(z_i-z_j\right)
\label{eq:ODE}
\end{equation}
for $i=1,2,3, \ldots, N$. In the expression $z_i$ is the position of each ion, $\omega_0$ the trap frequency along the axial direction, $k$ the Coulomb's constant, $q$ the charge of the electron, and $m$ is the mass of $\mathrm{^{40}Ca^+}$. To simulate this, we calculate the trajectory of each ion by considering a time-dependent trap potential and Coulomb interaction with all other ions. For initial states, the equilibrium positions of each ion are solved using the Newton method. The equilibrium positions of the nine-ion chain at the beginning are shown in Fig.~\ref{fig:equilibrium_positions}. Then, by simulating the trap potential with the finite element method, the temporal functions of the trap frequency and the trap center can be obtained. With the programmed half-speed shuttling function, the positions of all the ions during the transport are shown in Fig.~\ref{fig:Shuttling function and ions positions}. The expected relative $\bar{n}$ from transport-induced coherent excitation of nine motional modes along axial direction at the end of the half-speed shuttling process are shown in Fig.~\ref{fig:Relative amplitudes of 9 motional modes along axial direction}, showing the COM mode is dominantly excited in our simulation.

\section{Sideband cooling for the nine-ion chain}
\label{app:sideband}

In this section, we discuss the cooling scheme in this experiment and how thermal heating and coherent excitation affect carrier Rabi flopping with nine ions.
\subsection{Cooling scheme}

\begin{figure}[!t]
    \centering
    \includegraphics[width = 0.48\textwidth]{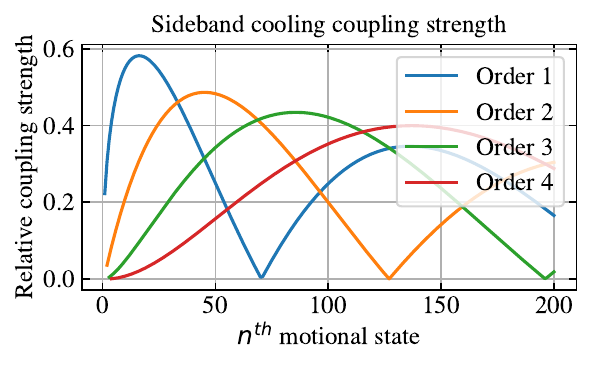}
    \captionsetup{justification=raggedright,singlelinecheck=false}
    \caption{Relative coupling strength on the carrier and the first three orders of the red sideband for $\eta$\,=\,0.23.}
    \label{fig:Sideband_cooling_coupling_efficiency}
\end{figure}

We use $\left|4^2 S_{1 / 2}, m_J=+1 / 2\right\rangle \leftrightarrow \left|3^2 D_{5 / 2}, m_J=-3 / 2\right\rangle$ for optical pumping, $\left|4^2 S_{1 / 2}, m_J=-1 / 2\right\rangle \leftrightarrow \left|3^2 D_{5 / 2}, m_J=-1 / 2\right\rangle$ to probe the carrier transition, and $\left|4^2 S_{1 / 2}, m_J=-1 / 2\right\rangle \leftrightarrow \left|3^2 D_{5 / 2}, m_J=-5 / 2\right\rangle$ for sideband cooling.
Since the axial COM frequency is as low as 179\,kHz, the Doppler cooling limit is around 43 quanta, placing the ions outside the Lamb-Dicke regime before starting the sideband cooling. So we need to consider the change of red sideband coupling strength at different motional states
\begin{equation}
\begin{aligned}
\Omega_{n-s, n} & =\Omega_0\left|\left\langle n-s\left|e^{i \eta\left(a+a^{\dagger}\right)}\right| n\right\rangle\right| \\
& =\Omega_0 e^{-\eta^2 / 2} \eta^{s} \sqrt{\frac{(n-s) !}{n !}} L_{(n-s)}^{s}\left(\eta^2\right),
\label{equ:red sideband coupling strength at different motional states}
\end{aligned}
\end{equation}
in which $s\,(>0)$ is the order of the red sideband~\cite{leibfried2003quantum}. As shown in Fig.~\ref{fig:Sideband_cooling_coupling_efficiency}, after Doppler cooling, the first-order sideband cooling may have inefficient coupling in a certain range of motional states. Therefore, we first conduct sideband cooling on the second-order red sideband of the three lowest frequency modes and then the first-order red sideband for all the modes. We also use additional cycles for sideband cooling on the COM mode since we found this is the dominant mode for thermal motions by scanning the motional spectrum. All these sideband cooling cycles are performed relative to carrier transition $\left|4^2 S_{1 / 2}, m_J=-1 / 2\right\rangle \leftrightarrow \left|3^2 D_{5 / 2}, m_J=-5 / 2\right\rangle$. On resonance, optical pumping cycles are inserted between sideband-cooling cycles and performed at the end.

\subsection{Spectrum scan}

\begin{figure}[!t]
    \centering
    \includegraphics[width = 0.48\textwidth]{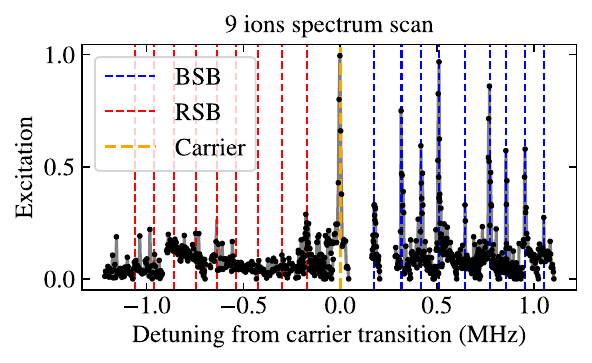}
    \captionsetup{justification=raggedright,singlelinecheck=false}
    \caption{Motional spectrum after sideband cooling covering the carrier, all 9 RSBs, and BSBs.}
    \label{fig:Spectrum scan covering all motional modes}
\end{figure}

After the sideband cooling and optical pumping cycles, we run a full spectrum scan covering the frequency range of all the first-order motional modes along the axial direction. As shown in Fig.~\ref{fig:Spectrum scan covering all motional modes}, even though additional cycles are added for the COM mode, it dominates the motional excitation of the crystal after sideband cooling, similar to what has been observed in previous research~\cite{wu2023continuous}.

\subsection{Carrier Rabi flop after sideband cooling}

\begin{figure}[!t]
    \centering
    \includegraphics[width = 0.48\textwidth]{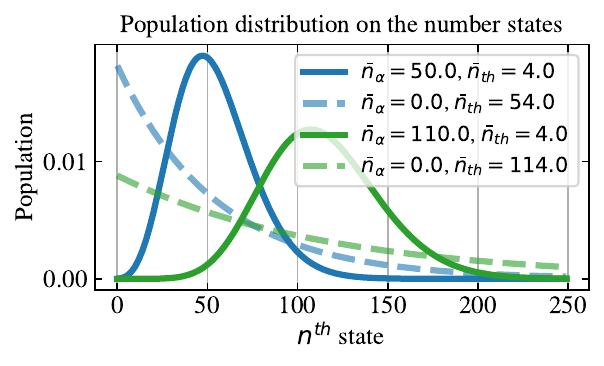}
    \captionsetup{justification=raggedright,singlelinecheck=false}
    \caption{Population distribution over number states under thermal heating, coherent excitation, and both.}
    \label{fig:Population distribution on the number states}
\end{figure}

The carrier Rabi frequency is modulated according to motional state occupation. In theory, all motional modes need to be taken into consideration since they have different Lamb-Dicke parameters. However, in this experiment, both the spectrum scan and the shuttling simulation show the COM mode dominating the motional excitation. Therefore, we only consider the COM mode for fitting carrier Rabi flops to reduce computational complexity. The carrier Rabi frequency is~\cite{cahill1969ordered}
\begin{equation}
\begin{aligned}
\Omega_{n n} & = \Omega \, e^{-|\alpha|^2 / 2} L_n\left(|\alpha|^2\right),
\end{aligned}
\end{equation}
in which $\alpha=i \tilde \eta e^{i \nu t}$, $\nu$ is the trap frequency along the axial direction, $\tilde{\eta}$ is the effective Lamb-Dicke parameter, which for the COM mode equals $\eta / \sqrt{N} \approx 0.077$ with ${N}$ being the number of ions. So the dependence of carrier Rabi frequency on the Lamb-Dicke parameter is
\begin{equation}
\begin{aligned}
\Omega_{n n} & = \Omega \, e^{-\tilde \eta^2 / 2} L_n\left(\tilde \eta^2\right) \\
&\approx \Omega[1 - (n + \frac{1}{2})\tilde \eta^2 + (\frac{1}{8}+\frac{n}{4}+\frac{n^2}{4})\tilde \eta^4 \\
&- (\frac{1}{48}+\frac{n}{18}+\frac{n^2}{24}+\frac{n^3}{36})\tilde \eta^6+...]
\end{aligned}
\end{equation}

Here, we expand the carrier Rabi frequency to the sixth order because of the low trap frequency and high motional excitation so that the ions are out of the Lamb-Dicke regime.

While noise sources like surface noise and technical noise will lead to a thermal state, fast shuttling causes coherent excitation.
Fig.~\ref{fig:Population distribution on the number states} shows that for coherent excitation, motional states will distribute around the $\bar{n}_\alpha=|\alpha|^2$ number state, while thermal heating leads to a Boltzmann distribution of the number states. In the presence of both mechanisms, the phonon probability follows a convolution of coherent and thermal distributions, which has a much wider distribution over different number states~\cite{walther2012controlling, ziesel2013quantum}.

\bibliography{multiplexing}

\end{document}